\documentclass[12pt]{article}
\usepackage{geometry}
\usepackage{graphicx}
\usepackage{amssymb,amsmath}
\usepackage{verbatim}

\newcommand{\beq}{\begin{equation}}
\newcommand{\eeq}{\end{equation}}

\newcommand{\bra}[1]{\langle{#1}|}
\newcommand{\ket}[1]{|{#1}\rangle}
\newcommand{\ps}[1]{\hspace{-5ex}{\phantom{\an{\hat A_{\mathrm{w}}}}}_{#1}}
\newcommand{\an}[1]{\left\langle{#1}\right\rangle}

\title{On the Weak Measurement of Velocity\\ in Bohmian
  Mechanics\footnote{This paper is dedicated to J\"urg Fr\"ohlich on the
    occasion of one of his birthdays.}}

\author{
Detlef D\"urr\footnote{Mathematisches Institut,
     Ludwig-Maximilians-Universit\"at, Theresienstr.~39, 80333
     M\"unchen, Germany. Email: duerr@mathematik.uni-muenchen.de}, Sheldon Goldstein\footnote{Departments of Mathematics, Physics and
     Philosophy, Hill Center, Rutgers, The State University of New  
     Jersey, 110 Frelinghuysen Road, Piscataway, NJ 08854-8019, USA.
     E-mail: oldstein@math.rutgers.edu},\\
 and Nino Zangh\`\i\footnote{Dipartimento di Fisica dell'Universit\`a
     di Genova and INFN sezione di Genova, Via Dodecaneso 33, 16146
     Genova, Italy. E-mail: zanghi@ge.infn.it} }
\date{August 13, 2008}
\begin{document}
\maketitle
\begin{abstract}
  In a recent article \cite{Wiseman}, Wiseman has proposed the use of
  so-called weak measurements for the determination of the velocity of a
  quantum particle at a given position, and has shown that according to
  quantum mechanics the result of such a procedure is the Bohmian velocity
  of the particle.  Although Bohmian mechanics is empirically equivalent to
  variants based on velocity formulas different from the Bohmian one, and
  although it has been proven that the velocity in Bohmian mechanics is not
  measurable, we argue here for the somewhat paradoxical conclusion that
  Wiseman's weak measurement procedure indeed constitutes a genuine
  measurement of velocity in Bohmian mechanics. We reconcile the apparent
  contradictions and  elaborate on some of the different senses of
  measurement at play here. 
\end{abstract}
\section{Introduction}
According to the uncertainty principle, it is impossible to simultaneously
measure both the position and the velocity of a quantum particle, at least
not to arbitrary accuracy. The basic reason for this limitation is that in
quantum mechanics a measurement of the position of a particle to a given
accuracy produces a corresponding narrowing of its wave function and hence
a corresponding increase in the uncertainty about its velocity. Moreover,
this is as true of Bohmian mechanics, a version of quantum mechanics in
which a particle always has a velocity as well as a position, as it is of
orthodox quantum theory, in which it does not. This suggests that in order
to measure the velocity of a Bohmian particle at a given position, it might
be good to exploit a measurement procedure that somehow does not
significantly affect the  wave function of the particle.

Such a procedure, a so-called {\it weak measurement,\/} has been developed
by Aharonov, Albert, and Vaidman \cite{AhaAlbVai88}. And Howard Wiseman has
indeed proposed in a recent article\cite{Wiseman} that weak measurements
be used to measure the velocity of a particle at a given position.

More precisely, Wiseman invokes  the theory of
weak measurements to provide an ``{\it operational definition} for the
velocity for a particle at position $x$'':
\beq 
{ v}({ x}) \equiv \lim_{\tau\to 0} \,\tau^{-1}\, {\rm
E}[{ x}_{\rm strong}(\tau) -  { x}_{\rm weak}|{ x}_{\rm
strong}(\tau) = { x}]. \label{1} 
\eeq
He then  observes that this quantity is precisely the velocity that defines
Bohmian mechanics \cite{Bohm52,Bellbook,qe,op}, and uses this fact to respond
to some objections that have been raised against it. 

In this formula $ { x}_{\rm weak}$ and ${ x}_{\rm strong}$ denote
respectively the results of a weak measurement of the position of the
particle at some time, say $t=0$, and a strong measurement of the position
a short time $\tau$ later. The expectation symbol E in the formula refers
to the average over a large ensemble of systems, all prepared in the same
initial state $\psi$ at time 0, and for all of which the result of a strong
measurement of position at time $\tau$, following the weak measurement at
time 0, is $x$.

A strong measurement of an observable $\hat A$ is just a standard quantum
measurement of the observable---one which collapses the wave function of the
system involved to the eigenstate of $\hat A$ corresponding to the
eigenvalue found in the measurement. The average of such values for a large
ensemble of systems in the state $\psi$ is, of course, $\langle \psi|\hat
A|\psi\rangle$. In contrast, a weak measurement of $\hat A$ (which will be
described in more detail in the next section) does not collapse the wave
function of the system, and in fact is such that the change in the wave
function that is produced by the procedure can be made arbitrarily small.
The price to be paid for this desirable feature is that very little
information about the system is obtained in a single such measurement, the
result found reflecting mainly the effect of noise introduced by the
procedure rather than any property of the system itself. Nonetheless, the
ensemble average for such a procedure is $\langle \psi|\hat A|\psi\rangle$,
just as for a strong measurement.

Weak measurements are most interesting when combined with post-selection:
Consider the subensemble for which, after the weak measurement, the system
is found in state $\varphi$ at time $\tau>0$. When $\tau=0+$, the average
over this subensemble of the result of the weak measurement at time 0 is of
course still $\langle \psi|\hat A|\psi\rangle$ when $\varphi=\psi$, and in
general is given by the so-called {\it weak value\/}
\cite{AhaAlbVai88}\footnote{Our usage here is that of Wiseman
  \cite{Wiseman}. It is a bit different from that of \cite{AhaAlbVai88},
  which refers to the ratio following ``Re'' in 
  equation   (\ref{w}), which  could be complex, as the weak value.}
 \begin{equation}
   \ps{\bra{\varphi}\!}\an{\hat
     A_{\mathrm{w}}}_{\ket{\psi}\phantom{}}=\,
\mathrm{Re}\frac{\langle {\varphi }|\hat A|{\psi }\rangle
}{\langle {\varphi }|{\psi }\rangle }. \label{w}
\end{equation}
For $\tau>0$ the value of the subensemble average of course involves the
unitary evolution operator $U(\tau)$ for time $\tau$ and is given by the
weak value
\begin{equation}
 \ps{\bra{\varphi}U(\tau)\!}\an{\hat
A_{\mathrm{w}}}_{\ket{\psi}\phantom{}}=\,
\mathrm{Re}\frac{\langle {\varphi }|{U(\tau)}\hat A|{\psi }\rangle
}{\langle {\varphi }|{U(\tau)}|{\psi }\rangle }. \label{2}
\end{equation}

In terms of this, the velocity definition (\ref{1}) becomes
\beq \label{wv} 
{ v}({ x}) = \lim_{\tau\to 0}
\,\tau^{-1}\left[ { x} - \mathrm{Re}\frac{\langle { x
}|{U(\tau)}\hat X|{\psi }\rangle }{\langle {x }|{U(\tau)}|{\psi
}\rangle }\right], 
\eeq
where $\hat X$ is the position operator of the particle.
One easily computes with $ U(\tau) =
\exp(-i \hat H\tau/\hbar)$ and $\hat{H} = {\hat{ p}^2}/{2m} +
V(\hat{ x})$ that (\ref{wv}) becomes 
\beq \label{3} 
{ v}({
x})=v^{\psi}(x)\equiv\frac{{ j^{\psi}} ({ x})}{|\psi({ x})|\,^2} 
\eeq 
with
\beq\label{j} { j^{\psi}} ({ x}) = ({\hbar}/{m}){\rm Im}\, \psi({
  x}) \mathbf{\nabla}\psi({ x}), 
\eeq 
the usual quantum flux.  (\ref{3}) is the expression for the Bohmian
velocity:
\begin{comment}We used for ease of comparison the notation in
  \cite{Wiseman}, but from now on we shall indicate--when the expressions
  depend on the wave function--that dependence by a super-index
  $\psi$.

We remark that the Bohmian trajectories are integral
curves along the velocity vector field, i.e. the equation defining the
Bohmian paths is next to Schr\"odinger's equation for the wave function
\cite{DGZ:Cushing}:
\end{comment} 
\begin{equation}\label{Bohm}
{ \dot X}(t)= v^\psi({ X}(t),t)=\frac{{ j}\,^\psi ({
X}(t);t)}{|\psi(t,{ X}(t))|\,^2}\,.
\end{equation}
This equation, together with Schr\"odinger's equation for the wave
function, is the defining dynamical equation of Bohmian mechanics for a
single particle, with a similar equation for the Bohmian mechanics of a
many-particle system.    

Wiseman does not claim, either in the above quotation or anywhere else in
his article, that his weak measurement procedure, providing an
``operational definition for the velocity,''  actually {\it measures\/} the
Bohmian velocity. There is a good reason for this: There are variants of
Bohmian mechanics based on velocity formulas different from (\ref{3},
\ref{j}) that yield theories empirically equivalent to Bohmian
mechanics. The existence of a procedure to measure the velocity would seem
to contradict this  empirical equivalence.

We elaborate. Recall the quantum continuity equation
\begin{equation}\label{quantumflux}
\partial_t |\psi(x,t)|\,^2 =-\mbox{div}\, { j}\,^\psi
({ x},t)\,. 
\end{equation}
From this equation ${ j}\,^\psi$ is not uniquely defined: A divergence free
vector can be added without affecting the continuity equation. In
\cite{Deotto} some ``physically reasonable'' additions are discussed,
giving rise to (empirically equivalent) variants of Bohmian mechanics which
have different velocity fields (\ref{Bohm}) with $ j\,^\psi$ replaced by
the new $ j\,'\,^\psi$. With respect to this Wiseman \cite{Wiseman}
describes his findings as follows:

\begin{quote}``...a particular ${\bf j}$ is
  singled out if one requires that ${\bf j}$ be determined {\em
    experimentally} as a {\em weak value}, using a technique that would
  make sense to a physicist with no knowledge of quantum mechanics. This
  ``naively observable'' ${\bf j}$ seems the most natural way to define
  ${\bf j}$ operationally. Moreover, I show that this operationally defined
  ${\bf j}$ equals the standard ${\bf j}$, so, assuming $\dot{\bf x}={\bf
    j}/P$ one obtains the dynamics of BM. It follows that the possible
  Bohmian paths are naively observable from a large enough
  ensemble.''
\end{quote} 
Notice that Wiseman claims only that the Bohmian
paths (or Bohmian velocities) are ``naively observable,'' but not that they
are {\it genuinely\/} observable. He claims not that the current (and the
velocity associated with it) found in his procedure is ipso facto the
actual current, but only that it is ``the most natural way to define
  ${\bf j}$ operationally.'' In short,  Wiseman does not claim that his
  procedure, which we shall call a ``weak measurment of velocity,''
  provides a genuine measurement of velocity.

But, as we shall argue, it does---despite the apparent contradiction. In
more detail, we shall be concerned in this paper with the following statements:
\begin{itemize}
\item[(1)] A ``weak measurement of velocity'' in Bohmian mechanics is, in a
  reasonable sense, a {\em genuine}  measurement of velocity.
\item[(2)]The same thing is true for the variants of Bohmian mechanics based on a velocity formula different from the Bohmian one
  mentioned above.
\item[(3)] Bohmian mechanics and the variants referred to in (2) are
  empirically equivalent to each other---and to standard quantum mechanics.
  In particular, for all of them the result of a ``weak measurement of
  velocity'' is given by the Aharonov-Albert-Vaidman formula given above,
  and hence by the formula for velocity in Bohmian mechanics.
\item[(4)] It is impossible to measure the velocity in Bohmian mechanics
  \cite{op}. 
\end{itemize}
There is of course an obvious contradiction between the first three of
these statements. A genuine measurement of velocity must reveal the
velocity and hence could be used to empirically distinguish the theories
based on different velocity formulas. And if a theory is based on a
velocity formula different from the Bohmian one, a genuine velocity
measurement for the theory can't yield the Bohmian velocity.

At least one of these three statements must be false.  However (3) is well
established, and true. (The reason for this is basically that in Bohmian
mechanics and its variants the statistics for the results of
experiments---including weak measurements---are determined by the same
$|\Psi|^2$ probabilities as for orthodox quantum theory.) In Section
\ref{ba} we shall show that (1) is also correct. We shall do this by an
analysis that seems to apply to the variants of Bohmian mechanics referred
to in (2), as well as to Bohmian mechanics itself. Thus in Section \ref{ba}
it shall seem as if we establish (2) as well as (1). In Section
\ref{careful} we shall explain why the analysis in Section
\ref{ba} is in fact
incorrect for the alternatives to Bohmian mechanics, so that (2) is not
established by the analysis yielding (1)---a good thing since (2) is false.

We shall also examine, in Section \ref{bmcc}, the crucial condition
responsible for the success of Bohmian mechanics here, showing directly
that this condition indeed uniquely characterizes Bohmian mechanics.
Finally, in Section
\ref{imp}, we address the apparent contradiction between statements
(1) and (4). 

\section{Bohmian Analysis of Weak Measurement\\ of Velocity}\label{ba}
We now consider a Bohmian particle with wave function $\psi$ at time 0. We
model the measurement apparatus which measures weakly the position of the
Bohmian particle by a pointer and we denote the actual Bohmian pointer
position by $Y$. We denote by $X(t)$ the position of the Bohmian particle.
We measure $X=X(0)$ weakly at time $0$ and very shortly after that, at time
$\tau$, we perform a strong measurement of $X(\tau)$.

Let us spell out what this means. Let $\Phi=\Phi(y)$ be the wave function
of the apparatus in its ready state, with the pointer-variable $Y$ centered
at $Y=0$: $\Phi$ is a real wave packet of spread $\sigma$---i.e., such that
$\Phi(y)=\phi(y/\sigma)$ with $\phi$ fixed as $\sigma\to\infty$---for which
the expected value of $Y$ vanishes,\footnote{This condition on the initial
  apparatus state $\Phi$, assumed for our analysis of the weak measurement
  of velocity in Bohmian mechanics, is weaker than what is in general
  needed for the result of a weak measurement, with post-selection and
  averaging, to be given by (\ref{w}), namely that $\phi$ be real and even,
  or real and odd.}
\begin{equation}\label{center}
\int dy\, y\, |\Phi(y)|^2 = 0\,,
\end{equation}
for example,
\begin{equation}\label{Gauss}
\Phi(y)\sim e^{-\frac{y^2}{4\sigma^2}}\,. 
\end{equation}
\begin{comment}
  We do
not spell out the interaction between particle and apparatus but
we assume that there is one which does the job appropriately, i.e.
there exists a unitary (Schr\"odinger-) evolution of the system
coupled to the apparatus which correlates the apparatus states
with the system state as appropriate for a position measurement.
\end{comment}
The weak measurement begins at time 0 with an interaction between system
and apparatus that leads to the following instantaneous transition from
initial quantum state to final quantum state:
\begin{comment}
a standard von Neumann measurement implemented by the unitary transformation
\end{comment}
 \begin{equation}\label{weakmeasure} 
\psi(x)\Phi(y)\to \psi(x)\Phi(y-x)
\end{equation}
(corresponding in ket notation to 
\begin{equation}
\int dx\,\psi(x)\ket{x}\ket{\Phi}\to \int dx\,\psi(x)\ket{x}\ket{\Phi}_x
\end{equation}
where $\ket{\ }_x$ indicates translation by $x$).
Immediately after this, the pointer position is measured and recorded. When
the measured value is $Y$,
(up to normalization) the  system wave function after the measurement is,
by the projection postulate,
\begin{equation}\label{after}
 \psi_{0+}(x)= \psi_Y(x)\equiv \psi(x)\Phi(Y-x)\,.
\end{equation}
Note that since the result $Y$ is random,
with probability distribution 
\beq
\rho^Y (y) = \int dx |\psi (x)|^2 | \Phi(y-x)|^2\,,
\eeq
the system wave function $\psi_{0+}(x)$ is random as well. (In Bohmian
mechanics this wave function is called the {\em conditional wave function}
\cite{op}.)

For a standard von Neumann measurement of position, the spread $\sigma$ of
the apparatus wave function $\Phi$ is taken very small so that the wave
function (\ref{after}) is an approximate eigenstate of the position
operator concentrated near the value $x=Y$.  But in a weak measurement the
pointer wave function $\Phi (y)$ is very spread out ($\sigma$ is very
large), varying on the scale $\sigma$, whereas $\psi(x)$ varies on a scale
of order unity (near say 0). We thus should have from (\ref{after}) that
\beq
\psi_Y(x)\approx \Phi(Y)\psi(x)
\eeq
and that up to normalization
\begin{equation}\label{equal}
\psi_{0+}(x)\approx \psi(x)\,,
\end{equation}
with small error, of order $1/\sigma$. Although a single weak measurement does not measure the actual position of the particle, by averaging over a large sample of identical experiments one obtains information about  the mean value of position; we have, observing (\ref{Gauss})
%$$\int | \Phi(y-x)|^2 y dy = x$$,
 $$ \mathbb{E} (Y) \equiv \int   y \rho^Y (y) dy  = \int x \rho^X (x) dx \equiv
  \mathbb{E} (X)$$
where $\rho^X (x) = |\psi (x)|^2$.

The conditional probability  density of $Y$ given $X=x$   is
\begin{equation}\label{need}
\rho^Y(y\,|X=x) = \frac{\rho^{X,Y} (x,y)}{\rho^X(x)} = \frac{|\psi (x)|^2 | \Phi(y-x)|^2}{|\psi (x)|^2}
 = |\Phi (y-x)|^2 \,,
\end{equation}
and hence  in a weak
 measurement
 \begin{equation}\label{weakwave} \mathbb{E}(Y|X=x) \equiv \int y  \rho^Y(y\,|X=x) =x\,.
 \end{equation}

 The ``weak measurement of velocity'' is completed by performing at time
 $\tau$, on each member of the ensemble, a (strong) measurement of the
 position of the particle, and taking the conditional average indicated by
 (\ref{1}). Conditioning on the event that $X(\tau)=x$,
\begin{comment}
For simplicity, assume that the
 Hamiltonian governing the time evolution is $$ H= H_x + H_y \,,$$ where
 $H_x$ is the free Schr\"odinger Hamiltonian for the $x$- system and $H_y=0$.
 (This is because we may regard the result of the weak
   measurement at time 0 as having been recorded, or as otherwise stable
   and fixed, and for our purposes this amounts to  regarding the
   y-system between time 0 and time $ \tau$ as having a trivial wave function
   dynamics---and for the Bohmian version we would also have a trivial
   motion for $Y$ as well.)
\end{comment} 
the Bohmian version of this conditional average is
 \begin{equation}\label{B1}
 \lim_{\tau\to 0}\frac{1}{\tau}\mathbb{E}\left(x-Y|X(\tau)=x\right)=\lim_{\tau\to 0}\frac{1}{\tau}\left(x-\mathbb{E}\left(Y|X(\tau)=x\right)\right)\,.
 \end{equation}
 
To see how the Bohmian velocity comes in we note, writing $X$ for $X(0)$,
 that
\beq\label{weakalways} 
X(\tau)\approx X + v^{\psi_{0+}}\tau.
\eeq
 When   $\tau\to 0$ the error of this approximation
 is of smaller order than $\tau$.
By (\ref{equal}) we have that
\beq\label{vapprox}
v^{\psi_{0+}}\approx v^{\psi}(x)
\eeq
and hence that 
\beq\label{Xv}
X(\tau)\approx X + v^{\psi}(X)\tau\approx X + v^{\psi}(X(\tau))\tau\,.
\eeq
With this approximation we can identify the event 
$X(\tau)=x$ with the event $X=x-v^{\psi}(x)\tau$.
 Therefore by (\ref{weakwave})
\begin{equation}\label{ygivenxtau}
\mathbb{E}(Y|X(\tau)=x)\approx\mathbb{E}(Y|X=x-v^{\psi}(x)\tau)=x-v^{\psi}(x)\tau
\end{equation}
and thus for (\ref{B1}) we obtain
 \begin{equation}\label{weakcons}\lim_{\tau\to
 0}\frac{1}{\tau}\left(x-\mathbb{E}(Y|X(\tau)=x)
 \right)\approx v^\psi(x)\,,\end{equation}
\begin{comment} 
which is the (approximated) Bohmian version of (\ref{1}). That is,  as we said,   quantum mechanically given by
 (\ref{3}).

 We note that one may arrive at the same result also by computing directly the conditional probability
 \begin{equation}
 \rho^Y(y|X(\tau)=x) = \frac{\rho_\tau(x,y)}{\rho^X_\tau(x)}\,,
 \end{equation}
 where $\rho_\tau(x,y)$ is the time evolved probability distribution with initial condition
 $\rho_0(x,y)= |\psi (x)|^2 | \Phi(y-x)|^2$, and $ \rho^X_\tau(x)$ is the $x$-marginal distribution of $\rho_\tau(x,y)$.  The distribution
  $ \rho_\tau (x,y)$ solves the continuity equation
  (\ref{quantumflux}) for $j=v\rho$,
\begin{equation}\label{conteq}
\frac{\partial \rho_t(x,y) }{\partial t}   = - \mbox{div}_x (v \rho_t (x,y) )
\end{equation}
where $v= v^{\psi_t} (x)$, with the initial condition
 \begin{equation}
\rho_0 (x,y) = |\psi(x)|^2 |\Phi(y-x)|^2.
 \end{equation}
According to the standard Lagrangian method, the solution in the short interval $\tau$
is
 \begin{equation}\label{ynull}
   \rho_\tau   (x,y) \approx (1- \tau \mbox{div}_x (v  (x))  |\psi|^2\left(x-v(x)\tau\right)|\Phi|^2 \left ( y- [x-v(x)\tau]\right).
 \end{equation}
 and
 \begin{equation}
  \rho^X_\tau(x) \approx (1- \tau \mbox{div}_x (v  (x))
   |\psi|^2\left(x-v(x)\tau\right)\,.
   \end{equation}
 Thus
 \begin{equation}\label{ygivenxtau1}
  \rho^Y(y|X(\tau) = x) \approx |\Phi|^2 \left ( y- [x-v(x)\tau]\right)
 \end{equation}
 whence
 \begin{equation}
 \mathbb{E}(Y|X(\tau)=x)= \int y  \rho^Y(y|X(\tau)=x) dy \approx x-v(x)\tau \,.
 \end{equation}
\end{comment}
with the approximation becoming exact for a weak measurement, i.e., in the
limit $\sigma\to\infty$.

The details of this analysis, in particular equations (\ref{weakwave}),
(\ref{Xv}) and (\ref{weakcons}), show that in a ``weak measurement of
velocity'' in Bohmian mechanics the result of the averaging is $v^\psi$
precisely because the Bohmian particle had velocity $v^\psi$. We are thus
justified in asserting that for Bohmian mechanics this procedure of weak
measurement genuinely measures the Bohmian velocity. Now we ask: What is specifically
``Bohmian'' in formulas (\ref{weakwave}), (\ref{Xv}) and (\ref{weakcons})?
The answer, it would seem, is nothing. These formulas seem to hold for all
variants of Bohmian mechanics which have differentiable paths. But that is
incompatible with the (correct) weak measurement formula (\ref{wv}) that
yields the Bohmian velocity.
\begin{comment}
That is the puzzle again: (\ref{Xv}) holds
for all deterministic variants of Bohmian mechanics, but (\ref{weakcons})
says by the quantum mechanical computation that the velocity must be the
Bohmian one, hence the equality can't be right!  
\end{comment}
But where is the mistake? The answer is in fact not easy to find and lies
in scrutinizing more carefully weak measurements.

\section{A More Careful Analysis}\label{careful}
Now here is the catch:  Since (\ref{equal}) is only approximately satisfied
for $\sigma$ large, where $\sigma$ is the spread of $\Phi(y)$,  we cannot
in general dismiss the possibility that $v$ depends also on $Y$.  The $Y$ dependence of the conditional wave
function $\psi_{0+}(x)=  \psi(x)\Phi(Y-x)$ yields in general a $Y$ dependence of the induced
velocity field.\footnote{We note that the $Y$ dependence of the
velocity of the particle can  be simply computed from the velocity
formula which the Bohmian type theory provides given the wave
function of the entire system consisting of particle and
apparatus. There is thus no need to  introduce the conditional
wave function. The conditional wave function focuses however on
the source of the $Y$-dependence: The weak measurement does affect
the wave function--if only a tiny bit. That tiny bit changes the
velocity a tiny bit, having possibly  a big effect in weak
measurements.}  Therefore after the weak measurement we have
truthfully now $v^{\psi_{0+}}= v(x, Y)$ and thus for (\ref{weakalways}) we
have 
 \begin{equation}\label{weakalwayscorrect} 
   X(\tau)\approx  X + v(X(\tau),Y)\tau\,.
\end{equation}
as a better approximation than (\ref{Xv}), exact to order $\tau$,
and the left hand sides of (\ref{weakcons}) becomes
\begin{equation}\label{weakconscorrect}\lim_{\tau\to
 0}\frac{1}{\tau}\left(x-\mathbb{E}(Y|X=x-v(x,Y)\tau)\right)\,.\end{equation}
This expression is not anymore easy to handle and  in general it is
not equal to $v^\psi$. Before entering into more details we make the
trivial  observation that if the weak measurement were such that $
v(x, Y)=v^{\psi}(x)$, i.e., if 
\beq\label{cc}
v^{\psi\Phi_y}=v^{\psi}\,,
\eeq
where $\Phi_y(x)=\Phi(y-x)$, the analysis of Section \ref{ba} would be
correct and the ``weak measurement of velocity'' would indeed be a genuine
measurement of velocity yielding the result $v^\psi$. When (\ref{cc}) is
satisfied, as is the case for Bohmian mechanics, we have in fact by
(\ref{need}) that, to order $\tau$,
\beq\label{papprox}
 \rho^Y(y|X(\tau) =x) =\rho^Y(y|X =x-v^\psi(x)\tau)= |\Phi|^2 \left ( y- [x-v^\psi(x)\tau]\right)\,.
\eeq
 
With a variant of Bohmian mechanics, however, the velocity need not (in
fact, will not, see below) obey (\ref{cc}) exactly . Rather (\ref{cc}) will
hold only approximately, presumably with an error of order $1/\sigma$ since
$\Phi$ varies on scale $\sigma$. At first sight one might be inclined to
ignore this error. However one must be careful here, since in a weak
measurement a large quantity (here $Y$, of order $\sigma$) is averaged to
yield (because of near-perfect cancellation) a result of order unity. And a
small change in the probability distribution involved could lead to an
effect of order unity as well.

Indeed, an order $ 1/\sigma$  error in  (\ref{cc})
would be expected to yield an additional contribution to (\ref{papprox}) of
order $ \tau/\sigma$, yielding a contribution of order $\tau$ to
$\mathbb{E}(Y|X(\tau)=x)$ (since $Y$ is of order $\sigma$), and hence a
contribution of order unity in (\ref{B1}). These expectations are correct. 

Indeed, writing now $v^{\psi}_B$ for the velocity (\ref{3}) in
Bohmian mechanics, we have that (to order $\tau$)
\begin{eqnarray}
 \rho^Y(y\,|X(\tau) =x) &=& |\Phi|^2 \left ( y-
   [x-v^\psi_B(x)\tau]\right)\nonumber \\
&=& |\Phi|^2 \left ( y-
   [x-v^\psi(x)\tau]-[v^\psi(x)-v^\psi_B(x)]\tau]\right)\\
&\approx&
|\Phi|^2 \left ( y-
   [x-v^\psi(x)\tau]\right) +(v^\psi(x)-v^\psi_B(x))\tau \cdot\nabla_x|\Phi|^2 \left ( y-x\right)\nonumber
\end{eqnarray}
and since $\Phi(y)$ varies on scale $\sigma$ the second term on the last
line is of order $\tau/\sigma$. Thus whenever the velocity $v^\psi$ is
non-Bohmian, and in particular whenever (\ref{cc}) is not obeyed, the
conditional distributution of $Y$ given $X(\tau)$ is sufficiently affected
so as to vitiate the analysis of Section \ref{ba}. When the velocity is
non-Bohmian the ``weak measurement of velocity'' fails because of these
errors to be a genuine measurement of velocity. 

\section{Bohmian Mechanics and the Crucial Condition}\label{bmcc}
There is one issue that might still be puzzling here. We have seen that a
``weak measurement of velocity'' is in fact a genuine measurement of
velocity whenever the condition (\ref{cc}) is satisfied. And in this case
the velocity found must be the Bohmian velocity. This implies that
(\ref{cc}) can be satisfied only for Bohmian mechanics---that it
characterizes Bohmian mechanics among all of its variants. We shall now
provide a more general and precise formulation of this conclusion, as well
as a direct argument for it:
\begin{quotation}
Suppose $v^\psi$ defines a variant of Bohmian mechanics for which the
  condition 
\beq\label{ccc} 
v^{\psi\phi}=v^{\psi}
\eeq
 holds for all (differentiable) real-valued  functions $\phi$, or at
  least for a collection of such functions that is ``gradient-total,''
  i.e., such that at every point $x\in\mathbb{R}^3$, the collection of vectors
  $\nabla\phi(x)$ spans $\mathbb{R}^3$. Then  $v^\psi=v^\psi_B$.
\end{quotation}

A similar conclusion holds for a more general configuration space than
$\mathbb{R}^3$, for example for $\mathbb{R}^{3N}$. We note that for any
fixed (differentiable) real-valued function $\Phi$ that vanishes at
$\infty$ (but is not identically 0), the collection of functions
$\Phi_y(x)=\Phi(y-x),\ y\in \mathbb{R}^3,$ is gradient-total, since
otherwise there would be a direction in which $\Phi$ does not vary. [Note
also that for particles without spin, i.e., when $\psi$ is
complex-scalar-valued, then the condition (\ref{ccc}) amounts basically to
requiring that $v^\psi$ depend only on the phase $S$ of $\psi$ (arising
from the polar decomposition $\psi=Re^{iS/\hbar}$).]

Here is the proof: Recall that any variant of Bohmian mechanics, with
velocity $v^\psi$, must obey the continuity equation,  see (\ref{quantumflux}),
\begin{equation}\label{cont}
\partial_t |\psi(x,t)|\,^2 =-\mbox{div}\, (v^\psi
({ x},t)|\psi(x,t)|\,^2)\,. 
\end{equation}
Consider two such velocity functionals, $v^\psi_1$ and $v^\psi_2$. Since
they both are such that (\ref{cont}) is obeyed, we have that
\begin{equation}\label{dd}
\mbox{div}\, (j^\psi_1-j^\psi_2)=0,
\end{equation}
where $j^\psi_i=|\psi|^2v^\psi_i$. If they both also obey (\ref{ccc}),
we have that 
\begin{equation}
j^{\psi\phi}_i = |\phi|^2j^\psi_i.
\end{equation}
Then from (\ref{dd}), with $\psi$ replaced by  $\psi'=\psi\phi$, we have that
\beq
\mbox{div}\,[|\phi|^2 (j^\psi_1-j^\psi_2)]=0,
\eeq
and since 
\beq
\mbox{div}\,[|\phi|^2 (j^\psi_1-j^\psi_2)]=|\phi|^2\mbox{div}\,
(j^\psi_1-j^\psi_2) + \nabla|\phi|^2 \cdot(j^\psi_1-j^\psi_2) =
\nabla|\phi|^2 \cdot(j^\psi_1-j^\psi_2) 
\eeq
it follows that
\beq
\nabla|\phi|^2 \cdot(j^\psi_1-j^\psi_2)=0.
\eeq
Thus if the relevant collection of functions $\phi$ is gradient-total, we 
have that 
\beq
j^\psi_1=j^\psi_2
\eeq
and hence that
\beq
v^\psi_1=v^\psi_2.
\eeq
Since $v^\psi_B$ is a possible choice for $v^\psi$ the conclusion follows.
%\section{Some remarks}
%The reader might be puzzled by the fact that all the empirical

\section{The Impossibility of Measuring the Velocity in Bohmian Mechanics} \label{imp}

We have argued that by using weak measurements it is possible to measure
the velocity of a particle in Bohmian mechanics. We will now reconcile this
with the proven impossibility of measuring the velocity in Bohmian
mechanics \cite{op}, in the sense that no measurement procedure involving an
interaction between a particle and any sort of apparatus can yield a result
that conveys (with arbitrary precision) the velocity of the particle just
prior to the beginning of the procedure.

Let us first note that our weak measurement of velocity can, it seems, be
regarded as just such a procedure. It involves an ensemble of systems, each
with the same wave function. We can regard one member of the ensemble as
the special particle whose velocity is to be measured, with the other
members of the ensemble constituting (part of) the apparatus. The selection
of subensemble and averaging corresponding to (\ref{weakcons}), with $x$
the position of the special particle at time $\tau$, then conveys the
velocity of the special particle to arbitrary accuracy.

This procedure, however, is not of the sort contemplated in \cite{op}. It
is assumed there that neither the initial state $\Psi_{\rm app}$ of the
apparatus nor the interaction $H_{\rm int}$ between system and apparatus
depends on the initial state $\psi$ of the system. (This is reasonable
since the whole point of a measurement is to obtain information about a
system that would not otherwise be available.) Such measurements have been
called linear measurements, in contrast with the nonlinear measurements in
which either the initial state of the apparatus $\Psi_{\rm
  app}=\Psi^{\psi}$ or the interaction $H_{\rm int}= H^{\psi}$ depends upon
the state $\psi$ of the system.

The weak measurement of velocity discussed in this paper is clearly
nonlinear,\footnote{Were it possible to clone the wave function $\psi$,
  the ensemble could have been produced as part of an overall linear
  measurement, but cloning is not possible \cite{GC}.} and is thus not precluded by
the impossibility claim of \cite{op}. It should nonetheless be contrasted
with another nonlinear measurement of velocity for Bohmian mechanics:
perform a standard position measurement and plug the result into the
formula (\ref{3}) to obtain the corresponding
velocity. While it is difficult to take the latter ``measurement of
velocity'' seriously, and to regard it as anything more than cheating, the
weak measurement of velocity in Bohmian mechanics  is, as we have argued, a
genuine measurement of velocity, even though it is nonlinear.    

\section{Conclusion}
Measurement is a tricky and complicated business. Even when, as with
Bohmian mechanics and its variants, there is something to be measured, one
must be careful. With orthodox quantum theory and the ``measurement'' of
operators as observables, the situation is even more dangerous. We conclude
by quoting Bell \cite[page 166]{Bellbook} on this:
\begin{quote}     
A final moral concerns terminology. Why did such serious people take so
seriously axioms which now seem so arbitrary? I suspect that they were
misled by the pernicious misuse of the word `measurement' in contemporary
theory. This word very strongly suggests the ascertaining of some
preexisting property of some thing, any instrument involved playing a
purely passive role. Quantum experiments are just not like that, as we
learned especially from Bohr. The results have to be regarded as the joint
product of `system' and `apparatus,' the complete experimental set-up. But
the misuse of the word `measurement' makes it easy to forget this and then
to expect that the `results of measurements' should obey some simple logic
in which the apparatus is not mentioned. The resulting difficulties soon
show that any such logic is not ordinary logic. It is my impression that
the whole vast subject of `Quantum Logic' has arisen in this way from the
misuse of a word. I am convinced that the word `measurement' has now been
so abused that the field would be significantly advanced by banning its use
altogether, in favour for example of the word `experiment.'
\end{quote}

\bigskip\bigskip

\noindent \textit{Acknowledgments.} The work of S.~Goldstein was supported
in part by NSF Grant DMS-0504504. That of N. Zangh\`\i\ was supported in
part by INFN.


\begin{thebibliography}{10}

\bibitem{Wiseman} H.~M.~ Wiseman:
  Grounding Bohmian mechanics in weak values and Bayesianism, {\em New
    Journal of Physics} \textbf{9}, (2007) 165,
  doi:10.1088/1367-2630/9/6/165
\bibitem{AhaAlbVai88} Y. Aharonov, D. Z. Albert, and
  L. Vaidman: How the result of a measurement of a component of the spin of
  a spin-1/2 particle can turn out to be 100, {\em Phys. Rev. Lett.} {\bf
    60}, 1351--1354 (1988)
\bibitem{Bohm52} D.~Bohm: A suggested interpretation of the quantum
   theory in terms of ``hidden'' variables, Part I, {\em Phys.\ Rev.}\
   \textbf{85}, 166--179 (1952).\  D.~Bohm: A suggested interpretation
   of the quantum theory in terms of ``hidden'' variables, Part II,
   {\em Phys. Rev.}\ \textbf{85}, 180--193 (1952)
\bibitem{Bellbook} J.S.~Bell: \textit{Speakable and unspeakable in
     quantum mechanics} (Cambridge University Press, Cambridge, 1987)
\bibitem{qe} D.~D\"urr, S.~Goldstein, and N.~Zangh\`\i: Quantum
   equilibrium and the origin of absolute uncertainty, {\em J.\
     Statist.\ Phys.}\ \textbf{67}, 843--907 (1992), quant-ph/0308039
\bibitem{op} D.~D\"urr, S.~Goldstein, and N.~Zangh\`\i: Quantum
   equilibrium and the role of operators as observables in quantum
   theory, {\em J.\     Statist.\ Phys.}\ \textbf{116}, 959--1055
   (2004), quant-ph/0308038 
\bibitem{Deotto}E.~ Deotto and G.~C.~Ghirardi: Bohmian mechanics
revisited, {\em Found. Phys.} \textbf{28},  1--30 (1998)
\bibitem{GC}G.~C.~Ghirardi: Private communication, 1981
\end{thebibliography}
\end{document}